\documentclass[conference]{IEEEtran}
\usepackage{amssymb,amsthm,amsmath,epsfig,latexsym,graphicx,bm,nccmath}
\usepackage{mathtools}

\usepackage{epstopdf}
\usepackage{cite}
\usepackage{amsfonts}
\usepackage[ruled, lined, linesnumbered, commentsnumbered, longend]{algorithm2e}
\usepackage{textcomp}
\usepackage{xcolor}
\def\BibTeX{{\rm B\kern-.05em{\sc i\kern-.025em b}\kern-.08em
		T\kern-.1667em\lower.7ex\hbox{E}\kern-.125emX}}
\usepackage{caption}	
\usepackage{subcaption}
\usepackage[utf8x]{inputenc}
\usepackage{etoolbox}
\usepackage{booktabs}
\usepackage[top=0.71in,bottom=1in,left=0.625in,right=0.625in]{geometry}

\usepackage[skip=0pt]{caption}
\usepackage[font=footnotesize,skip=0pt]{subcaption}
\usepackage{easyReview} 

\makeatletter
\patchcmd{\@maketitle}
{\addvspace{0.5\baselineskip}\egroup}
{\addvspace{-1\baselineskip}\egroup}
{}
{}
\makeatother

\begin{document}
	
	\title{Closed-loop Uplink Radio Resource Management in CF-O-RAN Empowered 5G Aerial Corridor}
	
	\author{\IEEEauthorblockN{Manobendu Sarker\IEEEauthorrefmark{1}, Md. Zoheb Hassan\IEEEauthorrefmark{2}, and Xianbin Wang\IEEEauthorrefmark{3}} 
		\IEEEauthorblockA{\IEEEauthorrefmark{1}{Department of Computer and Software Engineering}, Polytechnique Montreal, Canada}
		\IEEEauthorblockA{\IEEEauthorrefmark{2}{Department of Electrical and Computer Engineering}, Universit\'{e} Laval, Qu\'{e}bec City, QC, Canada }
		\IEEEauthorblockA{\IEEEauthorrefmark{3}{Department of Electrical and Computer Engineering}, Western University, London, Canada}
	}
	
	\maketitle
		
	\begin{abstract}
		In this paper, we investigate the uplink (UL) radio resource management for 5G aerial corridors with an open-radio access network (O-RAN)-enabled cell-free (CF) massive multiple-input multiple-output (mMIMO) system. Our objective is to maximize the minimum spectral efficiency (SE) by jointly optimizing unmanned aerial vehicle (UAV)–open radio unit (O-RU) association and UL transmit power under quality-of-service (QoS) constraints. Owing to its NP-hard nature, the formulated problem is decomposed into two tractable sub-problems solved via alternating optimization (AO) using two computationally efficient algorithms. We then propose (i) a QoS-driven and multi-connectivity-enabled association algorithm incorporating UAV-centric and O-RU-centric criteria with targeted refinement for weak UAVs, and (ii) a bisection-guided fixed-point power control algorithm achieving global optimality with significantly reduced complexity, hosted as xApp at the near-real-time (near-RT) RAN intelligent controller (RIC) of O-RAN. Solving the resource-allocation problem requires global channel state information (CSI), which incurs substantial measurement and signaling overhead. To mitigate this, we leverage a channel knowledge map (CKM) within the O-RAN non-RT RIC to enable efficient environment-aware CSI inference. Simulation results show that the proposed framework achieves up to 440\% improvement in minimum SE, 100\% QoS satisfaction and fairness, while reducing runtime by up to 99.7\% compared to an interior point solver-based power allocation solution, thereby enabling O-RAN compliant real-time deployment. 
	\end{abstract}

	\vspace{-3mm}
	
	\section{Introduction}	
	
	Aerial corridors are structured three-dimensional routes designed for unmanned aerial vehicle (UAV) traffic operating beyond visual line of sight (BVLOS). They offer a scalable mechanism to concentrate flight risk away from populations and infrastructure while embedding connectivity into route design. Realizing this vision requires re-engineering terrestrial cellular systems to ensure reliable command-and-control (C2), localization, and data links through techniques such as uptilted antennas and coordinated interference management~\cite{bhuyan2022advances}. Coverage analyses across altitudes reveal that existing terrestrial macrocells and millimeter-wave deployments alone are insufficient, motivating new corridor planning that leverages multi-tier fifth generation (5G) connectivity and pre-planned handovers to ensure ubiquitous, low-latency service~\cite{cherif2021aerial}.
	
	To create aerial corridors with guaranteed coverage , cellular networks must collaboratively provide site-specific beam management and optimized UAV-to-base station associations~\cite{Tarafder2024}. The open radio access network (O-RAN) architecture is a compelling enabler, augmenting cellular capabilities with near-real-time control and cross-layer programmability via its RAN intelligent controller (RIC), particularly suited to dynamic UAV scenarios~\cite{KarimiBidhendi2024}. A recent work \cite{Bertizzolo2023} demonstrates that O-RAN-based closed-loop control jointly optimizing UAV location and transmission directionality delivers approximately 19\% uplink (UL) capacity gain while meeting high-definition video quality-of-service (QoS) on multi-cell testbeds. At the RAN-function level, the RIC orchestrates coordinated  association and resource allocation, improving energy efficiency and cooperative coverage for scalable aerial corridors~\cite{Li2024}.
	
	Although O-RAN-enabled coordination improves average system throughput \cite{Bertizzolo2023,Li2024}, aerial corridors demand stringent per-UAV QoS guarantees, as any loss of connectivity can jeopardize flight safety and C2 link reliability. The unique propagation characteristics of aerial UAVs present a significant hurdle; their strong (often unobstructed) line-of-sight (LoS) links to multiple ground open radio units (O-RUs) can create severe UL interference, necessitating joint UAV-O-RU association and power control. Furthermore, centralized coordination schemes typically require global channel state information (CSI), yet the standard O-RAN E2 interface is not designed to support the large-scale, low-latency CSI exchange required for real-time, network-wide optimization.
	
	In this paper, we address these intertwined challenges of guaranteed connectivity and interference management by proposing a novel framework that integrates the cell-free (CF) massive multiple-input multiple-output (mMIMO) concept within an O-RAN architecture. In CF mMIMO, users are served concurrently by multiple distributed access points, inherently providing macro-diversity and robust connectivity \cite{Bjornson2020_CF_survey}. To the best of our knowledge, the problem of ensuring multi-connectivity and mitigating interference for 5G aerial corridors in an O-RAN-enabled CF mMIMO system has not been addressed in the literature. The main contributions of this work are summarized as follows. 
	
	%
	%
	%
	
	\smallskip \noindent $\bullet$ We propose an O-RAN-enabled CF mMIMO framework integrating a channel knowledge map \cite{10430216} within the non-real-time (non-RT) RIC to enable scalable CSI acquisition and circumvent E2 interface limitations.
	
	\smallskip \noindent $\bullet$ We formulate a max-min spectral efficiency (SE) optimization problem jointly optimizing UAV-O-RU association and UL power under per-UAV QoS constraints, and develop an alternating optimization (AO) framework with two efficient algorithms: a QoS-driven association algorithm and a bisection-guided fixed-point power control method achieving global optimality.
	
	\smallskip \noindent $\bullet$ We demonstrate through extensive simulations that the proposed framework achieves substantial improvements in minimum SE, QoS satisfaction, and fairness while reducing runtime significantly compared to an interior point solver-based methods.
	
	\vspace{-4mm}
	\section{System Model}
	\label{sec:system_model}
	
\begin{figure}[tb]	
		\centering
		\includegraphics[scale=0.23]{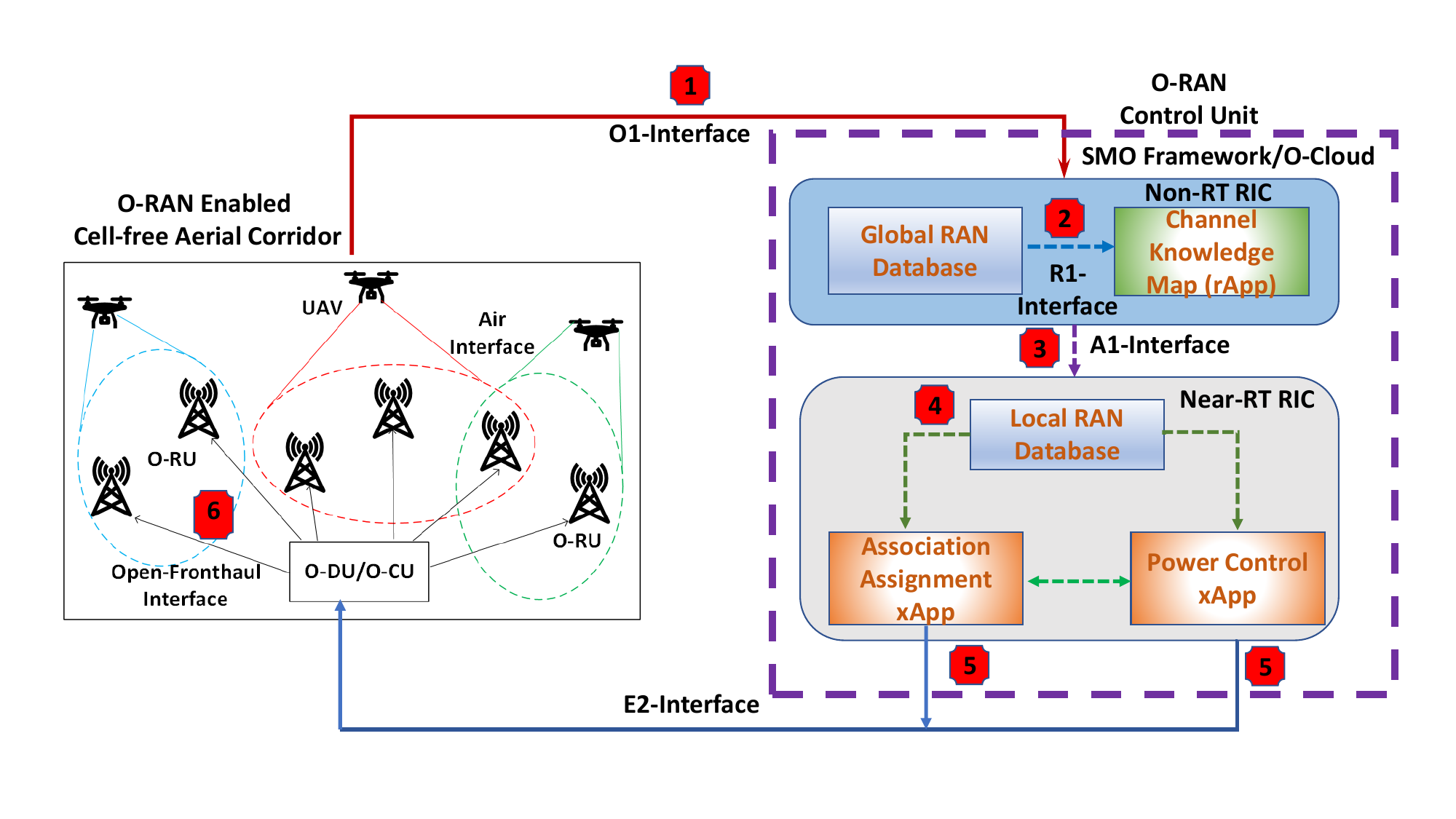}
		\caption{O-RAN-enabled CF mMIMO system for 5G aerial corridor.}
		\label{model}
	\end{figure}

	We consider an O-RAN-enabled CF mMIMO system supporting 5G aerial corridor, serving  $K$ single-antenna UAVs through $L$ geographically distributed O-RUs, shown in Fig. \ref{model}. Each O-RU $\ell \in \mathcal{L} = \{1, 2, \ldots, L\}$ is equipped with $N_\ell$ antennas and connected to an open distributed/centralized unit (O-DU/O-CU) via fronthaul links. The UAVs operate at altitudes ranging from ground level to several hundred meters and are indexed by $k \in \mathcal{K} = \{1, 2, \ldots, K\}$. Let $\mathcal{L}_k$ be the set of O-RUs that serve UAV $k$. We assume time-division duplex (TDD) operation, where the channel remains constant over a coherence block of $\tau_c$ symbols. Each coherence block is divided into $\tau_p$ symbols for UL pilot transmission and $\tau_c - \tau_p$ symbols for UL data transmission \cite{Bjornson2020_CF_survey}.
	
	\textbf{A Walk-through:} As illustrated in Fig.~\ref{model}, the proposed O-RAN-enabled framework comprises two core components: channel knowledge map (CKM) generation~\cite{10430216} and radio resource management (RRM) decision-making. The CKM serves as a repository of propagation statistics indexed by transmitter and receiver locations, thereby enhancing environmental awareness and reducing reliance on real-time CSI acquisition. By mitigating the challenges posed by high-dimensional channels and training overhead, CKM is envisioned as a key enabler for 6G networks demanding extreme capacity, ultra-low latency, and massive connectivity. CKM can be constructed using canonical interpolation methods, Kriging, kernel regression, matrix and tensor completion, deep learning, and environment model--assisted techniques~\cite{10430216}. In the proposed architecture, CKM is implemented as an rApp, a non-RT RIC application,  at the non-RT RIC, while real-time RRM is realized as an xApp, a near-RT RIC application, at the near-RT RIC. In what follows, we present the operational flow of the proposed framework: \textbf{(1)} The O-RAN network provides feedback and telemetry data to the service management and orchestration (SMO)/O-Cloud via the O1 interface, where it is stored in the global database (DB). \textbf{(2)} The global DB updates the CKM rApp through the R1 interface. \textbf{(3)} The CKM refreshes its information, and the Non-RT RIC supplies policy data (e.g., traffic information, RRM xApp configurations) and updated CSI to the local RAN DB via the A1 interface\footnote{RRM algorithms are implemented as xApps at the near-RT RIC. Because the E2 interface supports control signaling but not high-rate CSI exchange, CKM-derived CSI estimates (based on UAV positions obtained via E2~\cite{malik2024concept}) are delivered through the A1 interface instead.}. \textbf{(4)} Using Near-RT internal APIs, the local DB conveys control instructions and CSI to the RRM xApps. \textbf{(5)} The xApps compute RRM decisions and deliver them to the O-DU/O-CU through the E2 interface. \textbf{(6)} Finally, the O-DU/O-CU forwards these decisions to the O-RU via the open fronthaul, which transmits them to UAVs through the air-interface control channels.
	\vspace{-2mm}	  
	\subsection{Channel Model}
	\label{sec:channel_model}
	
	The channel between UAV $k$ and O-RU $\ell$ consists of: (i) large-scale fading capturing path loss, shadowing, and LoS probability; and (ii) small-scale Rician fading with dominant LoS component and multipath scattering \cite{Mozaffari2019_Tutorial}.
	
	\subsubsection{Large-Scale Fading}
	Following the 3GPP technical reports for aerial UAV scenarios \cite{3gpp.TR36.777,3gpp.TR38.901}, we adopt the enhanced urban macro-cell aerial (UMa-AV) propagation model that provides height-dependent LoS probability and distinct path loss expressions for LoS and non-LoS (NLoS) propagation conditions. Let $h_{\mathrm{UT},k}$ and $h_{\mathrm{O-RU},\ell}$ denote UAV and O-RU heights, and $d_{k\ell,2\mathrm{D}}$ the horizontal distance. The 3D distance is $d_{k\ell,3\mathrm{D}} = \sqrt{d_{k\ell,2\mathrm{D}}^2 + (h_{\mathrm{UT},k} - h_{\mathrm{O-RU},\ell})^2}$. LoS probability and path loss $\overline{\mathrm{PL}}_{k\ell}$ (in dB) follow distinct LoS/NLoS models from \cite{3gpp.TR38.901}. The large-scale coefficient is modeled as $\beta_{k\ell} = 10^{-(\overline{\mathrm{PL}}_{k\ell} + X_{\sigma})/10}$, where $X_{\sigma} \sim \mathcal{N}(0, \sigma_{\mathrm{sh}}^2)$ models log-normal shadowing. We assume $\beta_{k\ell}$ is known via long-term measurements \cite{Bjornson2020_CF_survey}.
	
	\subsubsection{Small-Scale Fading}
	The channel is modeled as spatially correlated Rician fading \cite{Geraci2019_CF_UAV}:
	\begin{equation}
		\scriptsize
		\mathbf{h}_{k\ell} = \sqrt{\beta_{k\ell}} \left( \sqrt{\frac{K_{k\ell}}{K_{k\ell} + 1}} \, \mathbf{a}_{k\ell}^{\mathrm{LoS}} + \sqrt{\frac{1}{K_{k\ell} + 1}} \, \mathbf{h}_{k\ell}^{\mathrm{scat}} \right) \in \mathbb{C}^{N_\ell},
	\end{equation}
	where $K_{k\ell}$ is the Rician K-factor, $\mathbf{a}_{k\ell}^{\mathrm{LoS}}$ is the array response with $\|\mathbf{a}_{k\ell}^{\mathrm{LoS}}\|^2 = N_\ell$, and $\mathbf{h}_{k\ell}^{\mathrm{scat}} \sim \mathcal{CN}(\mathbf{0}, \mathbf{R}_{k\ell})$ with $\mathrm{tr}(\mathbf{R}_{k\ell}) = N_\ell$. We use the mean-plus-deviation form $\mathbf{h}_{k\ell} = \overline{\mathbf{h}}_{k\ell} + \widetilde{\mathbf{h}}_{k\ell}$ \cite{Adhikary2013_Correlation}, where $\overline{\mathbf{h}}_{k\ell} = \sqrt{\beta_{k\ell} \frac{K_{k\ell}}{K_{k\ell} + 1}} \, \mathbf{a}_{k\ell}^{\mathrm{LoS}}$ with $\widetilde{\mathbf{h}}_{k\ell} \sim \mathcal{CN}\left(\mathbf{0}, \mathbf{C}_{k\ell}\right)$ and $\mathbf{C}_{k\ell} = \frac{\beta_{k\ell}}{K_{k\ell} + 1} \mathbf{R}_{k\ell}$.
	\vspace{-2mm}
	\subsection{Channel Estimation}
	\label{sec:channel_estimation}
	
	At the beginning of each coherence block, all $K$ UAVs simultaneously transmit pilot sequences to O-RUs. As in practical systems, we consider $K > \tau_p$, thereby pilot sequences must be reused among multiple UAVs, leading to pilot contamination. Considering MMSE channel estimation \cite{Kay1993_Estimation} is being used by O-DU/O-CU, the estimate $\widehat{\mathbf{h}}_{k \ell}$ and error $\widetilde{\mathbf{h}}_{k \ell}^{\mathrm{err}} = \mathbf{h}_{k \ell} - \widehat{\mathbf{h}}_{k \ell}$ are independent random variables with $\widehat{\mathbf{h}}_{k \ell} \sim \mathcal{CN}(\overline{\mathbf{h}}_{k \ell}, \widehat{\mathbf{C}}_{k \ell})$ and $\widetilde{\mathbf{h}}_{k \ell}^{\mathrm{err}} \sim \mathcal{CN}(\mathbf{0}, \mathbf{C}_{k \ell}^{\mathrm{err}})$, where $\widehat{\mathbf{C}}_{k \ell} = \mathbf{C}_{k \ell} - \mathbf{C}_{k \ell}^{\mathrm{err}}$ and $\mathbf{C}_{k \ell}^{\mathrm{err}} = \mathbf{C}_{k \ell} - \mathbf{C}_{k \ell} \boldsymbol{\Psi}_{k \ell}^{-1} \mathbf{C}_{k \ell}$ with $\boldsymbol{\Psi}_{k \ell} = \tau_p^2 \sum_{i \in \mathcal{P}_k} p_i^{\mathrm{p}} \mathbf{C}_{i \ell} + \tau_p \sigma^2 \mathbf{I}_{N_\ell}$. In $\boldsymbol{\Psi}_{k\ell}$, $\mathcal{P}_k$ is the set of UAVs sharing UAV $k$'s pilot and $p_k^{\mathrm{p}}$ is the pilot power, and $\sigma^2$ is the noise power. When $|\mathcal{P}_k| > 1$, $\widehat{\mathbf{h}}_{k\ell}$ is contaminated by LoS components of interfering UAVs.

	\subsection{Uplink Data Transmission and Spectral Efficiency}
	\label{sec:uplink}
	
	After pilot transmission, UAV $k$ transmits data with power $p_k^{\mathrm{u}} > 0$ over $\tau_c - \tau_p$ symbols. O-RU $\ell$ receives $\mathbf{y}_{\ell}^{\mathrm{u}} = \sum_{i=1}^K \sqrt{p_i^{\mathrm{u}}} \mathbf{h}_{i\ell} s_i + \mathbf{n}_{\ell}^{\mathrm{u}}$, where $s_k$ are unit-power data symbols and $\mathbf{n}_{\ell}^{\mathrm{u}} \sim \mathcal{CN}(\mathbf{0}, \sigma^2 \mathbf{I}_{N_\ell})$. Each O-RU applies L-MMSE combining $\mathbf{v}_{k\ell} = \left( \sum_{i=1}^K p_i^{\mathrm{u}} (\widehat{\mathbf{h}}_{i\ell} \widehat{\mathbf{h}}_{i\ell}^H + \mathbf{C}_{i\ell}^{\mathrm{err}}) + \sigma^2 \mathbf{I}_{N_\ell} \right)^{-1} \widehat{\mathbf{h}}_{k\ell}$ to detect UAV $k$ \cite{Bjornson2020_CF_survey}. The serving O-RUs $\mathcal{L}_k$ forward soft estimates to the CPU, which combines them with $\{\alpha_{k\ell}\}$, where $\alpha_{k\ell} = a_{k\ell}\sqrt{\beta_{k\ell}}$ are maximal ratio combining weights~\cite{Bjornson2020_CF_survey}. Using the use-and-then-forget bound \cite{Hassibi2003_Training}, the UL SE of UAV $k$ is $\mathrm{SE}_k = (1 - \tau_p/\tau_c) \log_2(1 + \Gamma_k)$, where $\Gamma_k$ is the signal-to-interference-plus-noise ratio (SINR), given in \eqref{eq:sinr_detailed}.
	
	\begin{figure*}[t]
		\footnotesize
		\begin{equation}
			\Gamma_k = \frac{\left|\sqrt{p_k^{\mathrm{u}}} \sum_{\ell \in \mathcal{L}_k} \alpha_{k\ell} \mathbb{E}\left[(\mathbf{v}_{k\ell})^H \mathbf{h}_{k\ell}\right]\right|^2}{p_k^{\mathrm{u}} \sum_{\ell \in \mathcal{L}_k} \alpha_{k\ell}^2 \left( \mathbb{E}\left[\left|(\mathbf{v}_{k\ell})^H \mathbf{h}_{k\ell}\right|^2\right] - \left|\mathbb{E}\left[(\mathbf{v}_{k\ell})^H \mathbf{h}_{k\ell}\right]\right|^2 \right) + \sum_{i \neq k} p_i^{\mathrm{u}} \sum_{\ell \in \mathcal{L}_k} \alpha_{k\ell}^2 \mathbb{E}\left[\left|(\mathbf{v}_{k\ell})^H \mathbf{h}_{i\ell}\right|^2\right] + \sigma^2 \sum_{\ell \in \mathcal{L}_k} \alpha_{k\ell}^2 \mathbb{E}\left[\|\mathbf{v}_{k\ell}\|^2\right]},
			\label{eq:sinr_detailed}
		\end{equation}
		\vspace{-6pt}
		\noindent\rule{\textwidth}{1pt}
	\end{figure*}
	\textbf{Remark.} SE calculation is required to finalize RRM decisions within the Near-RT RIC of O-RAN. Although the channel is estimated via CKM in our framework, the channel model and estimation are included here for the SE calculation model.
	\vspace{-2mm}
	\section{Problem Formulation}
	Our objective is to jointly optimize the UAV-O-RU association assignment and UL data transmit power allocation for maximizing the minimum SE performance while satisfying scalability and QoS requirements. To this end, we define an association matrix $\boldsymbol{A} \in \mathbb{R}^{K \times L}$ where $a_{k \ell} = 1$ if UAV $k$ associates with O-RU $\ell$, and $0$ otherwise. Thus, $\mathcal{L}_k = \{\ell: a_{k \ell} = 1\}$ with $|\mathcal{L}_k| = \sum_{\ell=1}^L a_{k \ell}$. Using these definitions, the SINR in (\ref{eq:sinr_detailed}) is reformulated as (\ref{eq:sinr_detailed_up}). Thus, the joint optimization problem is expressed as follows:
	\begin{figure*}[t]
		\footnotesize
		\begin{equation}
			\Gamma_k = \frac{\left|\sqrt{p_k^{\mathrm{u}}} \sum_{\ell =1}^L a_{k \ell} \alpha_{k\ell} \mathbb{E}\left[(\mathbf{v}_{k\ell})^H \mathbf{h}_{k\ell}\right]\right|^2}{p_k^{\mathrm{u}} \sum_{\ell =1}^L a_{k \ell} \alpha_{k\ell}^2 \left( \mathbb{E}\left[\left|(\mathbf{v}_{k\ell})^H \mathbf{h}_{k\ell}\right|^2\right] - \left|\mathbb{E}\left[(\mathbf{v}_{k\ell})^H \mathbf{h}_{k\ell}\right]\right|^2 \right) + \sum_{i \neq k} p_i^{\mathrm{u}} \sum_{\ell =1}^L a_{k \ell} \alpha_{k\ell}^2 \mathbb{E}\left[\left|(\mathbf{v}_{k\ell})^H \mathbf{h}_{i\ell}\right|^2\right] + \sigma^2 \sum_{\ell =1}^L a_{k \ell} \alpha_{k\ell}^2 \mathbb{E}\left[\|\mathbf{v}_{k\ell}\|^2\right]},
			\label{eq:sinr_detailed_up}
		\end{equation}
		\vspace{-6pt}
		\noindent\rule{\textwidth}{1pt}
	\end{figure*}

	\begin{subequations}
		\begin{alignat}{2}
			\textbf{P0:}\quad 
			\max_{\substack{\mathbf{A} \in \{0,1\},\\ \{p_k^{\mathrm{u}}\} \in [0,p^{\max}]}} \quad &
			\ \min_{k \in \mathcal{K}} \quad \text{SE}_k, 
			\label{eq:objective_unified} \\
			\text{s.t.}\quad 
			& \sum_{\ell \in \mathcal{L}} a_{k\ell} \geq 1, 
			&& \forall k \in \mathcal{K},  
			\label{c:ue_connectivity_unified} \\
			& \sum_{k \in \mathcal{K}} a_{k\ell} \leq \tau_p, 
			&& \forall l \in \mathcal{L}, 
			\label{c:ap_user_limit_unified} \\
			& \text{SE}_{k} \geq \text{SE}_k^{\min}, 
			&& \forall k \in \mathcal{K}. 
			\label{c:min_rate_unified}
		\end{alignat}
	\end{subequations}
	
	\noindent
	In \textbf{P0}, constraint~\eqref{c:ue_connectivity_unified} ensures every UAV connects to at least one O-RU for ensuring service coverage. Constraint~\eqref{c:ap_user_limit_unified} restricts each O-RU to serve at most  $\tau_p$  UAVs, dictated by pilot orthogonality, computational capacity, and scheduling overhead~\cite{sarker2023access}. Lastly, constraint~\eqref{c:min_rate_unified} enforces the minimum SE requirement  $\text{SE}_k^{\min}$  for QoS guarantees. However, \textbf{P0} is a mixed-integer nonlinear programming (MINLP) with binary association $\{a_{k\ell}\}$ and continuous power $\{p_k^{\mathrm{u}}\}$ variables. Because the objective~\eqref{eq:objective_unified} and minimum SE constraint~\eqref{c:min_rate_unified} both involve non-convex $\log_2(1 + \Gamma_k)$ functions, where $\Gamma_k$ depends nonlinearly on power and association. This renders \textbf{P0} NP-hard and computationally intractable.
	\vspace{-2mm}
	\section{Proposed Solution}
	\label{Proposed Solution}
	To address the NP-hardness in \textbf{P0}, we adopt a decoupling approach where the original problem is split into two sub-problems. Each sub-problem focuses on addressing a specific aspect, such as UAV-O-RU association assignment and UL data transmit power allocation.
	
	\subsection{UAV-O-RU Association Assignment}
	With fixed UL data transmit power $p_k^{\mathrm{u}}$, the UAV-O-RU association assignment problem is formulated as follows:
	\begin{subequations}
		\begin{alignat}{2}
			\textbf{P1:}\quad 
			\max_{\mathbf{A} \in \{0,1\}} \quad &
			\ \ \min_{k \in \mathcal{K}} \quad \text{SE}_k, \label{P1_obj}\\
			\text{s.t.}\quad &  \eqref{c:ue_connectivity_unified}\text{--}\eqref{c:ap_user_limit_unified}.
		\end{alignat}
	\end{subequations}
	Since the SINR expression is intricately linked to $\mathbf{A}$ through both channel combining and interference coupling, \textbf{P1} constitutes a MINLP. As finding the global optimum entails prohibitive computational complexity on the order of $\mathcal{O}(2^{KL})$, we develop an efficient heuristic approach to obtain a near-optimal solution, as detailed below.

	\subsubsection{Proposed UAV-O-RU Association Assignment Scheme}
	\label{Proposed UE-AP Association Assignment Scheme}
	
	To efficiently address the association problem, we propose a three-stage procedure for constructing a feasible, high-quality association matrix, summarized in Algorithm~\ref{alg:Association}.
	
	\smallskip \noindent $\bullet$ \textbf{Stage 1: UAV-centric initialization (Lines 3--7):} Each UAV \(k\) initially connects to its strongest O-RU \(\ell\) based on large-scale fading coefficients, ensuring universal connectivity under the O-RU capacity constraint~\eqref{c:ap_user_limit_unified}, thereby  satisfying constraint~\eqref{c:ue_connectivity_unified}.
	
	\smallskip \noindent $\bullet$ \textbf{Stage 2: O-RU-centric load balancing (Lines 8--14):} O-RUs with remaining capacity associate with their \(n_{\text{top}}\) strongest UAVs, where each O-RU \(\ell\) admits \(n_{\text{assign}}(\ell) = \min\!\left\{n_{\text{top}}, \tau_p - \sum_{k=1}^{K} a_{k\ell}\right\}\). This improves SE and balances load by leveraging unused capacity.
	
	\smallskip \noindent $\bullet$ \textbf{Stage 3: QoS-driven refinement (Lines 15--28):} UAVs violating QoS (\(\text{SE}_k < \text{SE}_k^{\min}\)) are identified and iteratively connected to additional candidate O-RUs (up to \(\lceil L/2 \rceil\)) in descending order of \(\{\beta_{k\ell}\}\), provided capacity is available. This stage ensures fairness by recovering QoS violations.

	\paragraph{Computational Complexity Analysis}
	\textbf{Stage~1} requires \(\mathcal{O}(KL \log L)\) operations as each UAV selects its strongest O-RU. Next, \textbf{Stage~2} sorts \(K\) UAVs per O-RU, yielding \(\mathcal{O}(LK \log K)\), and finally, \textbf{Stage~3} examines at most \(\lceil L/2 \rceil\) candidates for each QoS-violating UAV, with worst-case cost \(\mathcal{O}(|\mathcal{U}|L \log L)\). Hence, the total complexity is 
	\(\mathcal{O}(KL \log L + LK \log K + |\mathcal{U}|L \log L) \approx \mathcal{O}(KL)\), 
	since typically \(|\mathcal{U}| \ll K\).
	
	\begin{algorithm}
		\scriptsize
		\caption{Adaptive Joint UAV and O-RU-centric based Association Assignment Scheme}
		\label{alg:Association}
		
		\KwIn{$K$, $L$, $\tau_p$, $\boldsymbol{\beta}$,  $\text{SE}_k^{\min}$, $n_{\text{top}} = 3$}
		
		\KwOut{$\mathbf{A}^*$}
		
		\textbf{Initialize:} $\mathbf{A} \gets \mathbf{0}_{K \times L}$;

		\For{$k \gets 1$ \KwTo $K$}{
			$\ell^* \gets \arg\max_\ell \beta_{k\ell}$\;
			\If{$\sum_{k'=1}^K a_{k'\ell^*} < \tau_p $}{
				$a_{k\ell^*}^{*} \gets 1$\;
			}
		}
		
		\For{$\ell \gets 1$ \KwTo $L$}{
			Sort UAVs by descending $\beta_{k\ell}$: $\{k_1, k_2, \ldots, k_K\}$\;
			$n_{\text{assign}} \gets \min\left\{n_{\text{top}}, \tau_p - \sum_{k=1}^K a_{k\ell}^{*}\right\}$\;
			\For{$i \gets 1$ \KwTo $n_{\text{assign}}$}{
				$a_{k_i\ell}^{*} \gets 1$\;
			}
		}
		
		Compute $\{\Gamma_k\}$ and $\{\text{SE}_k\}$ with $\mathbf{A}^{*}$\;
		$\mathcal{U} \gets \{k : \text{SE}_k < \text{SE}_k^{\min}\}$\;
		
		\For{$k \in \mathcal{U}$}{
			$x \gets 1$\;
			\While{$\text{SE}_k < \text{SE}_k^{\min}$ \& $x \leq \lceil L/2 \rceil$}{
				$\mathcal{L}_k \gets \{\ell : a_{k\ell}^{*} = 1\}$\;
				Sort O-RUs by descending $\beta_{k\ell}$: $\mathcal{C} \gets \{\ell_1, \ldots, \ell_L\} \setminus \mathcal{L}_k$\;
				
				\If{$x \leq |\mathcal{C}|$}{
					$a_{k\mathcal{C}(x)}^{*} \gets 1$\;
					Recompute $\{\Gamma_k\}$ and update $\{\text{SE}_k\}$;
				}
				$x \gets x + 1$\;
			}
		}
		
	\end{algorithm}

	\subsection{UL Data Transmit Power Allocation}
	Given fixed UAV-O-RU association matrix $\mathbf{A}$, the UL data transmit power  allocation problem is formulated as follows:
	\begin{subequations}
		\begin{alignat}{2}
			\textbf{P2:}\quad  
			\max_{\{p_k^{\mathrm{u}}\} \in [0,p^{\max}]} \quad &
			\ \ \min_{k \in \mathcal{K}} \quad \text{SE}_k, \label{P2_obj}\\
			\text{s.t.}\quad &  \quad  \eqref{c:min_rate_unified}.
		\end{alignat}
	\end{subequations}
	Although the reformulated problem can be solved using an interior-point solver such as CVX \cite{Boyd2004_ConvexOpt}, the associated computational complexity is prohibitively high for near-RT RIC operation. To address the computational complexity of the CVX-based solver, we propose a \emph{bisection-guided fixed-point power control} (BG-FPPC) algorithm that achieves near-optimal max-min SINR performance with significantly reduced complexity, as described next. 
	
	\subsubsection{Proposed BG-FPPC Algorithm}
	\label{subsec:bg_fppc_algorithm}
	
	Our proposed BG-FPPC algorithm combines classical fixed-point iteration \cite{Foschini1993_power_control} with adaptive bisection to achieve near-optimal max-min SINR with reduced complexity. The complete procedure is summarized in Algorithm~\ref{alg:bg_fppc}.
	
	The algorithm starts by setting power of all UAVs at maximum power. Bisection SINR bounds are set as $\gamma_{\min}$ and $\gamma_{\max}$. The precomputed SINR coefficients $\{a_k, b_{ki}, c_k\}$  represent desired signal, inter-UAV interference, and noise, respectively.
	
	The algorithm employs two nested loops. An \textbf{outer bisection} (Lines~\ref{line:bisect_start}--\ref{line:bisect_end}) searches for the maximum feasible target SINR $\gamma^{\star}$ by iteratively testing the midpoint $\gamma_{\text{mid}} = (\gamma_{\min} + \gamma_{\max})/2$ (Line 3), while an \textbf{inner fixed-point iteration} (Lines~\ref{line:fp_start}--\ref{line:fp_end}) computes the minimum power required to achieve the current target via the update in Line~\ref{line:fp_update}. At each bisection step, feasibility is checked by verifying $\max_k p_k^u \leq p^{\max}$ (Line~\ref{line:feasibility_check}). The solution is stored if feasibility  is satisfied and $\gamma_{\min} $ is updated with the value of $ \gamma_{\text{mid}}$ (Lines~\ref{line:update_lower}--\ref{line:store_best}), otherwise it updates $\gamma_{\max} $ (Line~\ref{line:update_upper}). The process terminates when the relative gap between $\gamma_{\max}$ and $  \gamma_{\min}$ is below $ \epsilon_{\text{bisect}}$.
	
	The key novelty of Algorithm~\ref{alg:bg_fppc} is the {adaptive target SINR mechanism}. Rather than fixing $\gamma$ a priori as in standard Foschini-Miljanic approaches \cite{Foschini1993_power_control}, our algorithm systematically explores the feasible region to solve the max-min problem globally, avoiding local optima of gradient-based methods.
	
	\paragraph{Computational Complexity} Each fixed-point iteration incurs $\mathcal{O}(K^2)$ for interference computation. With $I_{\text{FP}} \approx 5$--$15$ inner iterations and $I_{\text{bisect}} = \mathcal{O}(\log(1/\epsilon_{\text{bisect}}))$ bisection steps, the total complexity is $\mathcal{O}(I_{\text{bisect}} \cdot I_{\text{FP}} \cdot K^2) = \mathcal{O}(K^2)$. Compared to CVX's $\mathcal{O}(K^{3.5})$ \cite{Boyd2004_ConvexOpt}, this represents a $\mathcal{O}(K^{1.5})$ reduction, confirmed by over $10\times$ runtime reduction in Section~\ref{sec:numerical_evaluation}.
	
	\begin{algorithm}[t]
		\caption{Bisection-Guided Fixed-Point Power Control (BG-FPPC)}
		\label{alg:bg_fppc}
		\small
		\KwIn{$\{a_k, b_{ki}, c_k\}$, $P_{\max}$, $\Gamma_{\text{init}}$, $\epsilon_{\text{bisect}} = 10^{-4}, \epsilon_{\text{FP}} = 10^{-3}$, $N_{\max}^{\text{FP}} = 20$}
		\KwOut{ $\mathbf{p}^{\star}$, $\gamma^{\star}$}
		
		\textbf{Initialize:} $\mathbf{p}_{\text{init}} \leftarrow p^{\max} \mathbf{1}$, $\mathbf{p}^{\star} \leftarrow \mathbf{p}_{\text{init}}$, $\gamma_{\min} \leftarrow 0$, $\gamma_{\max} \leftarrow 1.5 \cdot \max_k \Gamma_{\text{init},k}$, $\gamma^{\star} \leftarrow \min_k \Gamma_{\text{init},k}$\;
		
		\While{$(\gamma_{\max} - \gamma_{\min})/\gamma_{\max} > \epsilon_{\text{bisect}}$}{ \label{line:bisect_start}
			$\gamma_{\text{mid}} \leftarrow (\gamma_{\min} + \gamma_{\max})/2$ \;
			
			$\mathbf{p}^{(0)} \leftarrow p^{\max} \mathbf{1}$, $n \leftarrow 0$ \;
			\Repeat{$\|\mathbf{p}^{(n+1)} - \mathbf{p}^{(n)}\|_{\infty} < \epsilon_{\text{FP}} \cdot p^{\max}$ \textbf{or} $n = N_{\max}^{\text{FP}}$}{ \label{line:fp_start}
				\For{$k = 1, \ldots, K$}{ \label{line:fp_loop_start}
					$I_k \leftarrow \sum_{i \neq k} b_{ki} p_i^{(n)} + c_k$ , 
					$p_k^{(n+1)} \leftarrow \frac{\gamma_{\text{mid}}}{a_k} \cdot I_k$ \label{line:fp_update} \; 
				}
				$n \leftarrow n + 1$ \;
			} \label{line:fp_end}
			
			\label{line:feasibility_check} 	
			\eIf{$\max_k p_k^{(n)} \leq p^{\max}$}{ \label{line:feasibility_true}
				$\gamma_{\min} \leftarrow \gamma_{\text{mid}}$ \; \label{line:update_lower}
				$\mathbf{p}^{\star} \leftarrow \min(\mathbf{p}^{(n)}, p^{\max} \mathbf{1})$ \label{line:store_best},  
				$\gamma^{\star} \leftarrow \min_k \mathrm{SINR}_k(\mathbf{p}^{\star})$\; 
			}{
				$\gamma_{\max} \leftarrow \gamma_{\text{mid}}$ \; \label{line:update_upper}
			}
		} \label{line:bisect_end} \label{line:bisect_converge}
	\end{algorithm}
	\vspace{-2mm}
	\subsection{The Overall Solution to the Problem \textbf{P0}}
	\label{The Overall Solution to the Problem P0}
	
	We solve the coupled problem \textbf{P0} via AO, initialized with $p_k^{\mathrm{u}} = p^{\max}$. Each iteration alternates between: (i) solving for association $\mathbf{A}$ with fixed power (Algorithm~\ref{alg:Association}), and (ii) solving for power $\{p_k^{\mathrm{u}}\}$ with fixed $\mathbf{A}$ (Algorithm~\ref{alg:bg_fppc}). The process terminates when the objective~\eqref{eq:objective_unified} improvement falls below $\epsilon$ or iteration count reaches $I_{\max}$. Convergence is guaranteed via monotonic improvement of the bounded objective, typically within 3--5 iterations, with per-iteration complexity $\mathcal{O}(KL + K^2)$.
	\section{Numerical Evaluation}
	\label{sec:numerical_evaluation}
	This section evaluates the performance of the proposed UAV–O-RU association and UL transmit power allocation schemes in an O-RAN-enabled CF mMIMO network. Table~\ref{tab:sim_params} summarizes the key simulation parameters following 3GPP UMa-AV specifications \cite{3gpp.TR36.777,3gpp.TR38.901}. The AO algorithm uses $\epsilon = 0.001$ and $I_{\max}=15$ as convergence and iteration limits, respectively. UAV trajectories are assumed predetermined and fixed. Pilot sequences are randomly assigned to UAVs and transmitted at full power. All results are averaged over $500$ Monte Carlo realizations, each with independent O-RU and UAV positions, altitudes, and channel realizations. 
	
	\begin{table}[ht]
		\caption{Simulation Parameters}
		\label{tab:sim_params}
		\centering
		\small
		\begin{tabular}{lc}
			\toprule
			\textbf{Parameter} & \textbf{Value} \\
			\midrule
			Coverage area & $1 \times 1$ km$^2$ \\
			Number of O-RUs ($L$)  & $100$ \\
			Antennas per O-RU ($N_\ell$) &  $4$ \\
			UAV altitude range & $[50, 150]$ m \\
			Carrier frequency ($f_c$) & $2.6$ GHz \\
			Coherence block ($\tau_c$) / Pilot ($\tau_p$) & $200$ / $10$ symbols \\
			Rician $K$-factor range & $[0, 20]$ dB \\
			Shadow fading $\sigma_{\mathrm{sh}}$ (LoS / NLoS) & $4$ / $6$ dB \\
			Angular spread at O-RUs (mean) & $[5^\circ, 15^\circ]$ ($8^\circ$) \\
			Max UAV transmit power ($p^{\max}$) & $23$ dBm \\
			Noise PSD / Noise figure & $-174$ dBm/Hz / $9$ dB \\
			\bottomrule
		\end{tabular}
	\end{table}
	
	\subsubsection{Minimum SE Performance}
	\label{Comparison of Minimum SE Performance}
	
\begin{figure}[tb]	
		\centering
		\includegraphics[scale=0.18]{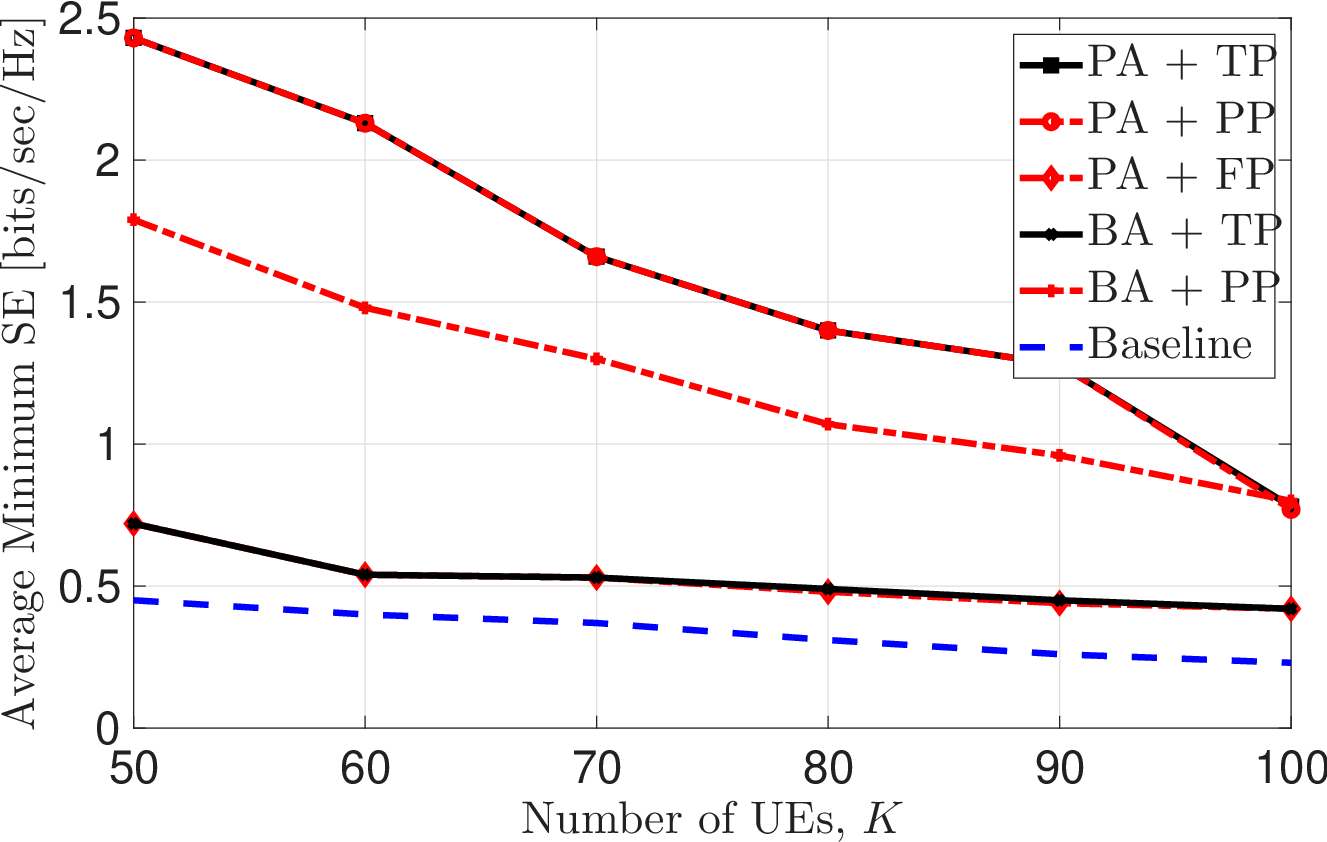}
		\caption{Average minimum SE performance of different schemes for varying UAV $K$ with $\mathrm{SE}^{\min}=1~\text{bit/s/Hz}$.}
		\label{minrate}
	\end{figure}

	Fig. \ref{minrate} compares the average minimum SE performance between six schemes combining different UAV-O-RU association and UL transmit power allocation strategies. For association, we consider: (a)~the baseline (BA) scheme from \cite{sarker2023access}, where both UAV and O-RU-centric associations are leveraged, and (b)~the proposed association (PA) in Algorithm~\ref{alg:Association}. For power allocation, we consider: (a)~full power transmission (FP), (b)~the proposed Algorithm~\ref{alg:bg_fppc} (PP), and (c)~a CVX-based solver applying bisection to problem~\textbf{P2} (TP). So, the six benchmarks are: (i)~`Baseline' (BA~+~FP), (ii)~`PA + FP', (iii)`BA~+~PP', (iv)~`BA~+~TP', (v)~`PA~+~PP', and (vi)~`PA~+~TP'. The AO framework from Section~\ref{The Overall Solution to the Problem P0} is applied only to schemes~(v) and~(vi), as the association schemes are computed independently of power allocation.  
	
	Fig.~\ref{minrate} demonstrates substantial improvements in the minimum SE performance achieved by the proposed schemes. The joint optimization of association and power allocation (PA~+~PP or PA~+~TP) attains up to a 440\% gain in minimum SE over the Baseline, highlighting the effectiveness of these two schemes. To isolate individual contributions, we note that the proposed association alone (PA) yields up to an 297.8\% improvement over the Baseline, owing to the QoS-driven refinement in Algorithm~\ref{alg:Association}, which dynamically assigns additional serving O-RUs to weak UAVs, a feature that is not included in~\cite{sarker2023access}. As for the proposed power allocation scheme (Algorithm \ref{alg:bg_fppc}), it improves the minimum SE performance by up to 82.6\% compared to the Baseline. Notably, the proposed BG-FPPC algorithm (PP) achieves the same minimum SE as the CVX-based solver (TP), confirming its global optimality. This equivalence arises because (a)~the outer bisection loop systematically searches for the maximum feasible target SINR $\gamma^{\star}$, and (b)~the inner fixed-point iteration precisely computes the minimum power required to meet each target, ensuring that no feasible solution is overlooked. As expected, the minimum SE decreases with increasing UAV density for all schemes due to higher inter-UAV interference. Interestingly, the proposed association schemes (PA~+~PP and PA~+~TP) exhibit steeper degradation than the baseline since they proactively assign more O-RUs to QoS-constrained UAVs, thereby increasing interference sensitivity under dense deployments. In contrast, the baseline association remains static and thus less responsive to variations in UAV density.
	
	\subsubsection{Success Rate Performance}
	\label{Comparison of Success Rate Performance}
	
\begin{figure}[tb]	
		\centering
		\includegraphics[scale=0.18]{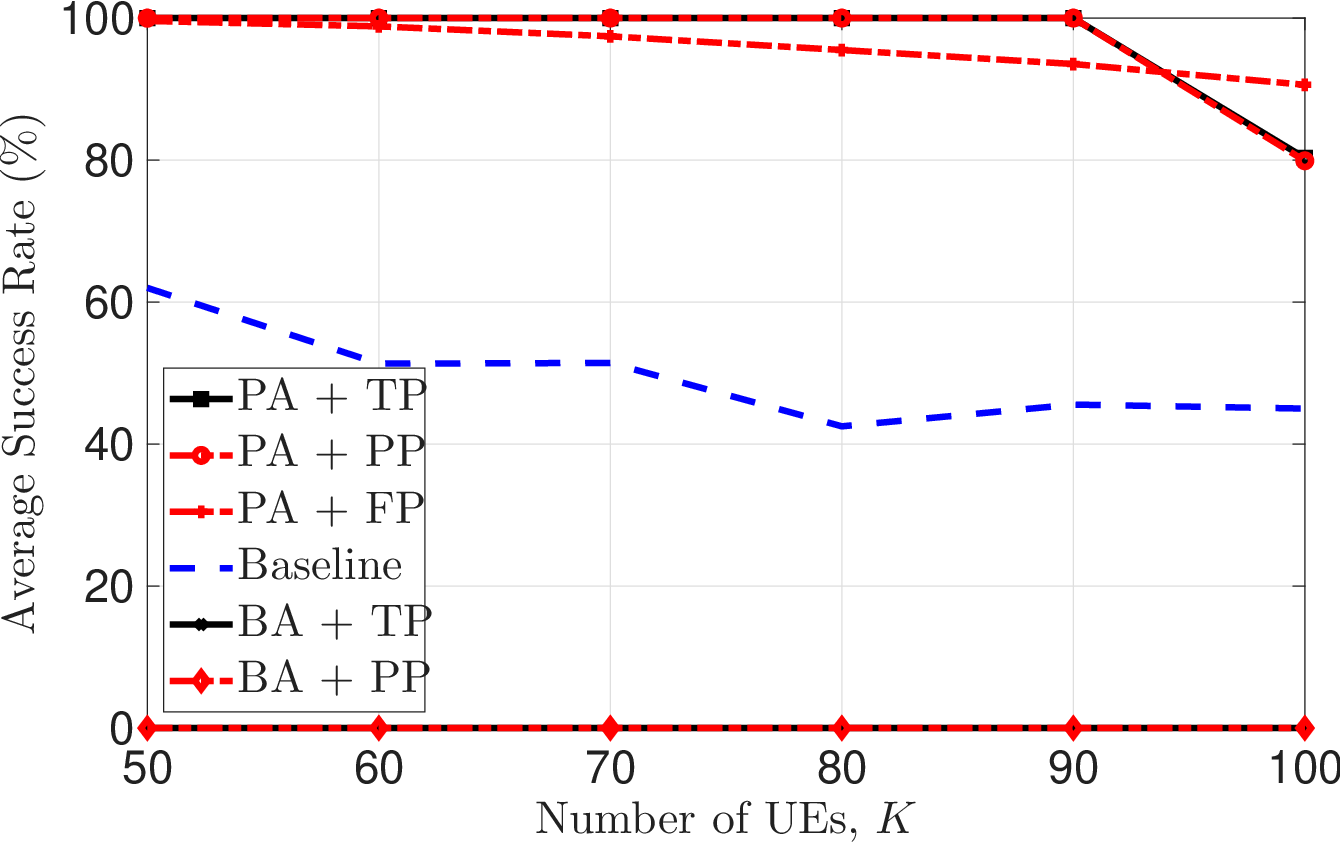}
		\caption{Average success rate performance of different schemes for varying UAV $K$ with $\mathrm{SE}^{\min}=1~\text{bit/s/Hz}$.}
		\label{success}
	\end{figure}

	Fig.~\ref{success} shows the average success rate performance, defined as the percentage of UAVs achieving an SE greater than the target threshold $\mathrm{SE}^{\min}$. Although the baseline association combined with power optimization (in BA~+~PP and BA~+~TP schemes) enhances the minimum SE (as shown in Fig.~\ref{minrate}), these schemes fail to meet $\mathrm{SE}^{\min}$ for any UAV, resulting in a zero success rate across all UAV densities. This limitation arises because the baseline association in~\cite{sarker2023access} does not incorporate QoS-aware O-RU selection, and power control alone cannot compensate for inadequate serving O-RU selection for weak UAVs. On the other hand, the proposed association scheme  achieves a 100\% success rate for $K \leq 90$ in PA~+~PP and PA~+~TP schemes, corresponding to up to a 135\% improvement over the Baseline. When $K = 100$, the success rate of PA~+~PP and PA~+~TP decreases to approximately 80\% due to resource scarcity under increased interference, whereas the Baseline exhibits minimal variation with density, consistent with the trend observed in Fig.~\ref{minrate}. It is important to note that PA~+~TP and PA~+~PP yield identical success rates across all UAV densities, indicating that both power allocation strategies (proposed and CVX-based) equally uphold the QoS guarantees ensured by Algorithm~\ref{alg:Association}. Interestingly, even PA~+~FP achieves success rates comparable to PA~+~PP and PA~+~TP for $K \leq 90$ (differences below 6.5\%), confirming that QoS-driven O-RU assignment in Algorithm~\ref{alg:Association} is the dominant factor in meeting $\mathrm{SE}^{\min}$, with power optimization providing marginal additional benefit. At $K = 100$, however, PA~+~FP provides a 13.4\% higher success rate than PA~+~PP and PA~+~TP because full power transmission offers greater robustness under resource scarcity.
	
	\subsubsection{Fairness Performance}
	\label{Comparison of Fairness Performance}
	
\begin{figure}[tb]	
		\centering
		\includegraphics[scale=0.18]{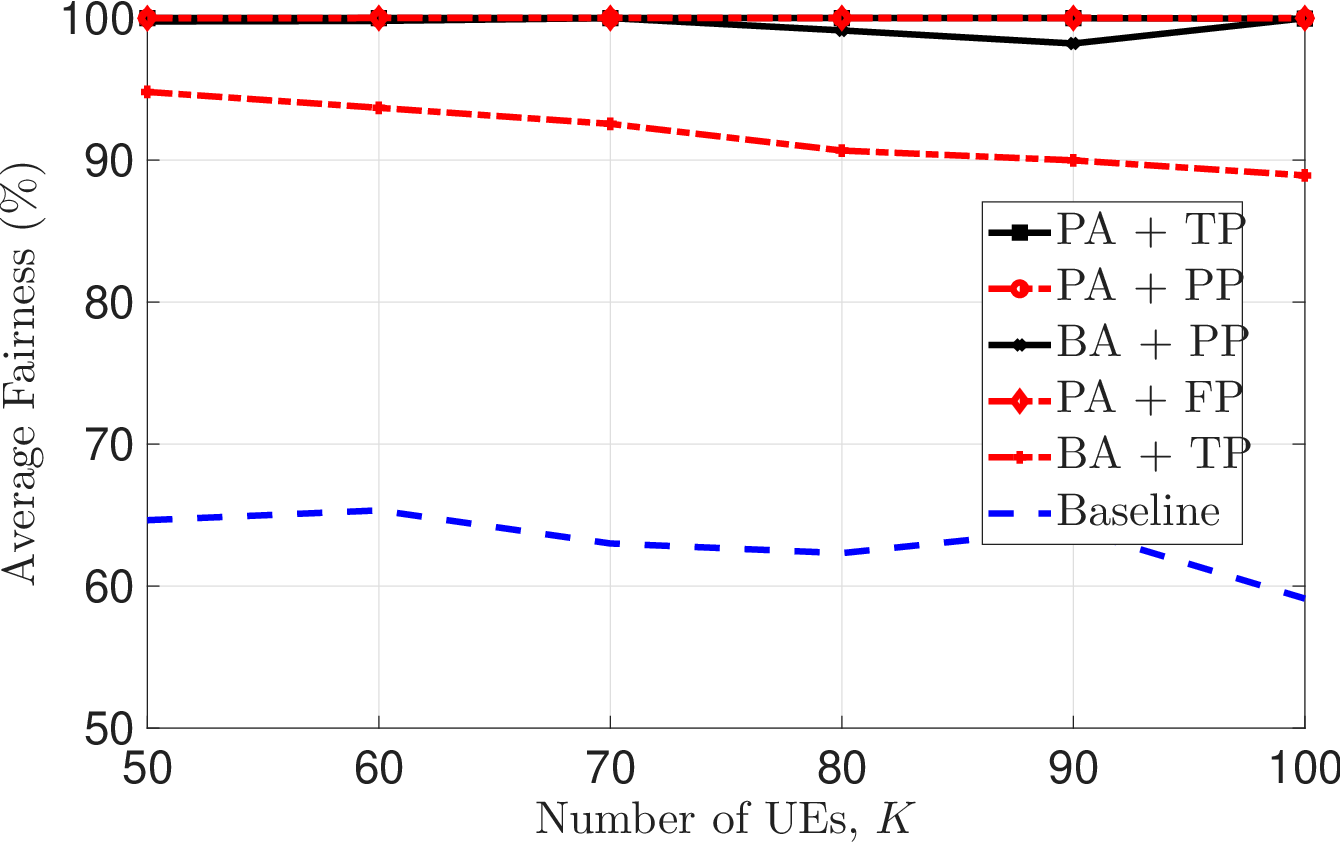}
		\caption{Average fairness performance of different schemes for varying UAV $K$ with $\mathrm{SE}^{\min}=1~\text{bit/s/Hz}$.}
		\label{fairness}
	\end{figure}

	Fig.~\ref{fairness} presents the average fairness performance, quantified in percentage using Jain's fairness index. All schemes employing power optimization (BA~+~PP, BA~+~TP, PA~+~PP, and PA~+~TP) achieve 100\% fairness, representing a 70\% improvement over the Baseline. This substantial gain arises because max-min power allocation inherently promotes equitable SE distribution by prioritizing power allocation to weak UAVs. In comparison, the proposed association alone (PA~+~FP) yields a 50.4\% improvement over the Baseline, confirming that QoS-driven O-RU assignment enhances fairness, though it achieves approximately 11\% lower fairness than power-optimized schemes. This gap occurs because, without power optimization, even effective O-RU association cannot fully equalize SE across UAVs with diverse channel conditions, establishing power control as the dominant fairness-enabling mechanism.   
	
	\subsubsection{Computational Complexity Performance}
	\label{Comparison of Computational Complexity Performance}
	
\begin{figure}[tb]	
		\centering
		\includegraphics[scale=0.18]{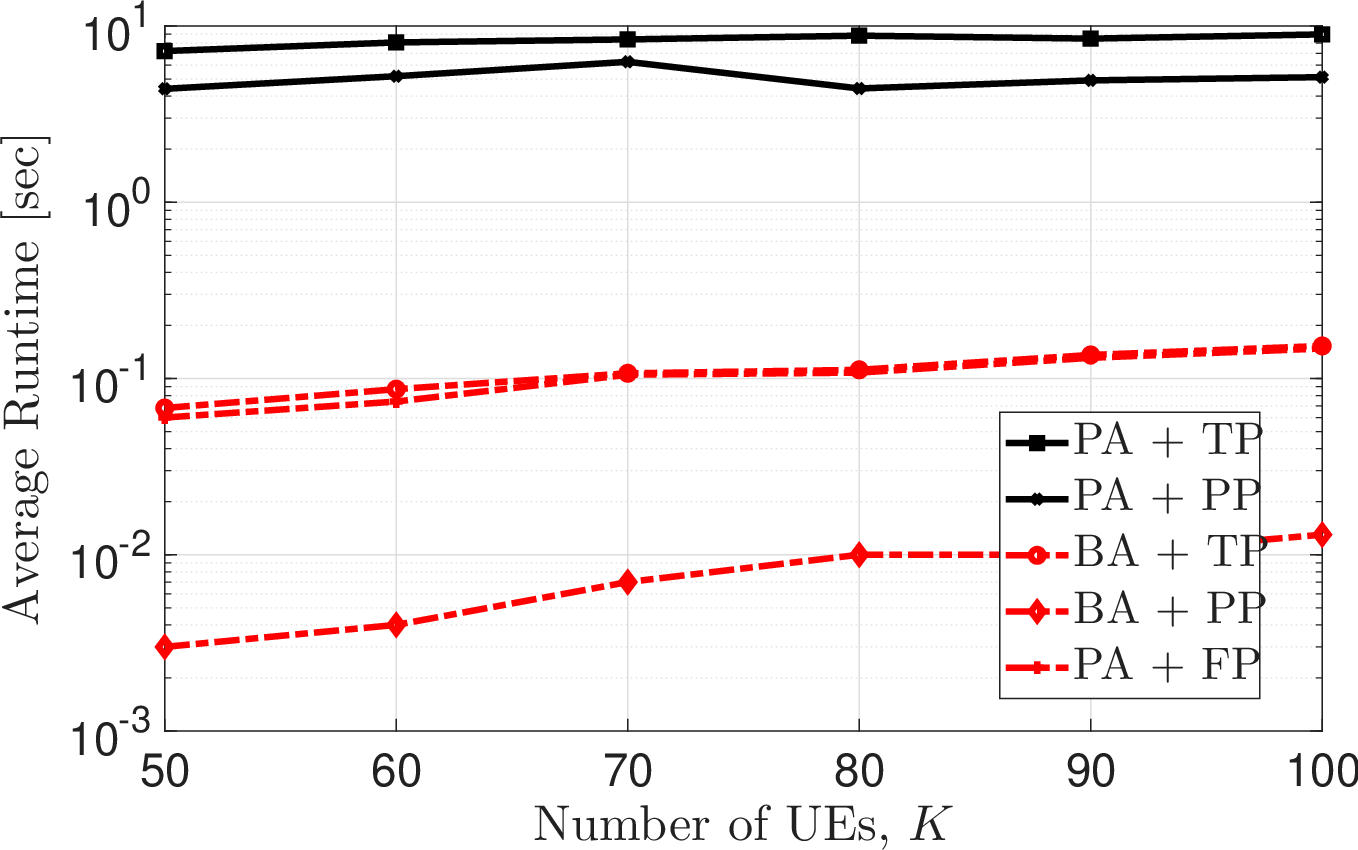}
		\caption{Average runtime of different schemes for varying UAV $K$ with $\mathrm{SE}^{\min}=1~\text{bit/s/Hz}$.}
		\label{runtime}
	\end{figure}

	Fig.~\ref{runtime} illustrates the average runtime for all schemes except the Baseline. The proposed BG-FPPC algorithm (PP) achieves up to 99.1\% runtime reduction compared to the CVX-based scheme (TP), with BA~+~PP outperforming BA~+~TP by up to 99.9\%. Both BA~+~PP and PA~+~PP schemes with proposed algorithm (Algorithm \ref{alg:bg_fppc}) achieve 10-150 ms duration, which is within the near-RT RIC time constraint runtime, enabling O-RAN compliant real-time deployment. Notably, PA~+~PP incurs only 3.4\% additional runtime compared to PA~+~FP, confirming that the AO framework for coordinating the proposed association and power allocation schemes introduces minimal computational overhead.
	
	\vspace{-3mm}
	\section{Conclusion}
	This paper presents a joint UAV-O-RU association and power allocation framework for O-RAN-enabled CF mMIMO systems supporting 5G aerial corridors. The key contributions include: (i)~an adaptive association algorithm that dynamically selects serving O-RUs by combining UAV and O-RU-centric clustering, followed by QoS-driven refinement for weak UAVs, (ii)~a computationally efficient power control method that integrates bisection search with fixed-point iteration achieving global optimality, and (iii)~an AO framework integrating both solutions. The framework adopts a CKM within the O-RAN non-RT RIC to mitigate CSI acquisition overhead. Simulation results demonstrated substantial enhancements in minimum SE, QoS compliance, fairness, and computational efficiency compared to baseline approaches. Our analysis reveals that association design is critical for QoS guarantees, while power control dominates fairness. The proposed schemes yield an optimal solution within sub-second runtime, suitable for O-RAN compliant real-time deployment. Future work will investigate joint UAV trajectory optimization to enhance coverage and SE in dynamic aerial networks.
	\vspace{-3mm}
	\bibliography{Ref_UAV}
	\bibliographystyle{ieeetr}
	
\end{document}